\documentclass[showpacs,preprintnumbers,amsmath,amssymb,nofootinbib,floatfix]{revtex4}

\usepackage[dvips]{graphicx}
\usepackage{dcolumn}
\usepackage{bm}

\def\mapgeq{\mathbin{\lower.3ex\hbox{$\buildrel>\over{\smash{\scriptstyle\sim}\vphantom{_x}}$}}}
\def\mapleq{\mathbin{\lower.3ex\hbox{$\buildrel<\over{\smash{\scriptstyle\sim}\vphantom{_x}}$}}}
\def\mapgeqeq{\mathbi{\lower.3ex\hbox{$\buildrel>\over{\smash{\scriptstyle\approx}\vphantom{_2}}$}}}
\def\mapleqeq{\mathbin{\lower.3ex\hbox{$\buildrel<\over{\smash{\scriptstyle\approx}\vphantom{_2}}$}}}
\def\Journal#1#2#3#4{{#1} {\bf #2} (#4) #3}
\def\MPL{Mod. Phys. Lett. A}

\def\NPB{Nucl. Phys. B}

\def\NPSUPPL{Nucl. Phys. Proc. Suppl.}
\def\PLB{{Phys. Lett.} B}

\def\PLBOLD{Phys. Lett.}
\def\PRL{Phys. Rev. Lett.}
\def\RMP{Rev. Mod. Phys.}
\def\PRD{Phys. Rev. D}

\def\PTP{Prog. Theor. Phys.}
\def\JHEP{JHEP}

\def\EPJ{Euro. Phys. J. C}

\def\JETPUSSR{Sov. Phys. JETP}

\def\ZETP{Zh. Eksp. Teor. Fiz.}

\def\IJMP{Int. J. Mod. Phys. A}

\def\JPG{J. Phys. G}

\def\NJP{New. J. Phys.}
\def\Erratum{Erratum-ibid}

\begin{document}

\preprint{TOKAI-HEP/TH-0505}

 \title{A New Type of Complex Neutrino Mass Texture and $\mu$-$\tau$ Symmetry}

\author{Ichiro Aizawa}
\email{4aspd001@keyaki.cc.u-tokai.ac.jp}

\author{Masaki Yasu\`{e}}%
\email{yasue@keyaki.cc.u-tokai.ac.jp}
\affiliation{\vspace{5mm}%
\sl Department of Physics, Tokai University,\\
1117 Kitakaname, Hiratsuka, Kanagawa 259-1292, Japan\\
}

\date{October, 2005}

\begin{abstract}
Relying upon the usefulness of the $\mu$-$\tau$ symmetry, we find a new type of neutrino mass texture with a single phase parameter $\delta$ that describes maximal atmospheric neutrino mixing and Dirac CP violation due to the presence of $\delta$.  The Majorana phase associated with the third massive neutrino turns out to be identical to the Dirac phase while other Majorana phases vanish.  The nonvanishing reactor neutrino mixing angle $\theta_{13}$ is induced by a $\mu$-$\tau$ symmetry breaking effect.  Flavor neutrino masses that supply the $\mu$-$\tau$ symmetry breaking terms become pure imaginary for $\delta=\pm \pi/2$, leading to maximal CP violation.  There is a parameter denoted by $\eta$, which is either ${\mathcal{O}}(\sqrt{\Delta m^2_\odot/\Delta m^2_{atm}})$ in the normal mass hierarchy, or ${\mathcal{O}}(\Delta m^2_\odot/\Delta m^2_{atm})$ in the inverted mass hierarchy.  In the inverted mass hierarchy, the contribution of ${\mathcal{O}}(\sin^2\theta_{13})$ is found to be significant and cannot be neglected. Our texture also leads to quasi degenerate neutrinos with masses of ${\mathcal{O}}(\sqrt{\Delta m^2_{atm}})$, which serves as the scale for the effective neutrino mass in $(\beta\beta)_{0\nu}$-decay.  This texture does not include $\eta$, and $\Delta m^2_\odot$ naturally arises from contributions of ${\mathcal{O}}(\sin^2\theta_{13})$ to give $\Delta m^2_\odot/\Delta m^2_{atm}\sim\sin^2\theta_{13}$, yielding the prediction of $\sin^2\theta_{13}={\mathcal{O}}(10^{-2})$.
\end{abstract}

\maketitle
\section{\label{sec:1}Introduction}
There have been growing theoretical interest in determining the structure of the complex neutrino mass matrix \cite{MassTextureCP,MassTextureCP1,MassTextureCP2,MassTextureCP3,MassTextureCP4} that induces leptonic CP violation \cite{CPViolation}.  The leptonic CP violation has two sources: one from a CP violating Dirac phase and the other from two CP violating Majorana phases \cite{CPViolationOrg} if neutrinos are Majorana particles.  It is then argued that the $\mu$-$\tau$ symmetry \cite{Nishiura,mu-tau,mu-tau1,mu-tau2} plays a crucial role in the physics of leptonic CP violation \cite{MassTextureCP3}.  In the standard parameterization of CP violation used by PDG \cite{PDG},  the PMNS unitary matrix \cite{PMNS} is given as $U^{PDG}_{PMNS}=U_\nu K$:
\begin{eqnarray}
U_\nu&=&\left( \begin{array}{ccc}
  c_{12}c_{13} &  s_{12}c_{13}&  s_{13}e^{-i\delta}\\
  -c_{23}s_{12}-s_{23}c_{12}s_{13}e^{i\delta}
                                 &  c_{23}c_{12}-s_{23}s_{12}s_{13}e^{i\delta}
                                 &  s_{23}c_{13}\\
  s_{23}s_{12}-c_{23}c_{12}s_{13}e^{i\delta}
                                 &  -s_{23}c_{12}-c_{23}s_{12}s_{13}e^{i\delta}
                                 & c_{23}c_{13}\\
\end{array} \right),
\nonumber \\
K &=& {\rm diag}(e^{i\beta_1}, e^{i\beta_2}, e^{i\beta_3}),
\label{Eq:U_nu}
\end{eqnarray}
for $c_{ij}=\cos\theta_{ij}$ and $s_{ij}=\sin\theta_{ij}$ ($i,j$=1,2,3), where $\theta_{ij}$ stands for the three neutrino mixing angles. The three flavor neutrinos $\nu_{e,\mu,\tau}$ are converted by the action of $U^{PDG}_{PMNS}$ into three massive neutrinos $\nu_{1,2,3}$.  The CP violating Dirac phase is denoted by $\delta$, while the Majorana CP violation phases are determined by two combinations of $\beta_{1,2,3}$ such as $\beta_i-\beta_3$ ($i$=1,2,3).  In this parameterization, the Dirac phase $\delta$ drops out if $s_{13}=0$, which is the solution required by the $\mu$-$\tau$ symmetry.  Therefore, Dirac CP violation depends on the presence of the $\mu$-$\tau$ symmetry breaking part.  These mixing angles are constrained by various experiments \cite{SK,Experiments} and are currently summarized as \cite{NuData}
\begin{eqnarray}
\sin ^2 \theta _{12}  = 0.314
{\footnotesize
\left( {1\begin{array}{*{20}c}
   { + 0.18}  \\
   { - 0.15}  \\
\end{array}} \right)}
,
\quad
\sin ^2 \theta _{23}  = 0.44
{\footnotesize
\left( {1\begin{array}{*{20}c}
   { + 0.41}  \\
   { - 0.22}  \\
\end{array}} \right),
}
\quad
\sin ^2 \theta _{13}  = 0.9 
{\footnotesize
{\begin{array}{*{20}c}
   { + 2.3}  \\
   { - 0.9}  \\
\end{array}} \times 10^{-2}.
}
\label{Eq:NuDataAngle}
\end{eqnarray}
Neutrino masses enter in the solar and atmospheric neutrino mass squared differences defined by $\Delta m^2_\odot = m^2_2-m^2_1$ $(>0)$ \cite{PositiveSolor} and $\Delta m^2_{atm} = \vert m^2_3-(m^2_1+m^2_2)/2\vert$, which are observed to be:
\begin{eqnarray}
\Delta m^2_\odot = 7.92\left( 1\pm0.09\right) \times 10^{-5}~{\rm eV}^2,
\quad
\Delta m^2_{atm} = 2.4
{\footnotesize
\left( {1\begin{array}{*{20}c}
   { + 0.21}  \\
   { - 0.26}  \\
\end{array}} \right)
} \times 10^{-3}~{\rm eV}^2.
\label{Eq:NuDataMass}
\end{eqnarray}

Let us introduce a neutrino mass matrix $M_\nu$ parameterized by\footnote{It is understood that the charged leptons and neutrinos are rotated, if necessary, to give diagonal charged-current interactions and to define the flavor neutrinos of $\nu_e$, $\nu_\mu$ and $\nu_\tau$.}
\begin{eqnarray}
&& M_\nu = \left( {\begin{array}{*{20}c}
	M_{ee} & M_{e\mu} & M_{e\tau}  \\
	M_{e\mu} & M_{\mu\mu} & M_{\mu\tau}  \\
	M_{e\tau} & M_{\mu\tau} & M_{\tau\tau}  \\
\end{array}} \right).
\label{Eq:NuMatrixEntries}
\end{eqnarray}
It is then convenient to divide $M_\nu$ into the $\mu$-$\tau$ symmetric part $M_{sym}$ and symmetry breaking part $M_b$ \cite{MassTextureCP3,MassTextureCP4} expressed in terms of $M^{(\pm)}_{e\mu} = (M_{e\mu} \pm (-\sigma M_{e\tau}))/2$ and $M^{(\pm)}_{\mu\mu} = (M_{\mu\mu} \pm M_{\tau\tau})/2$ for $\sigma=\pm 1$:
\begin{eqnarray}
&&
M_\nu = M_{sym} + M_b
\label{Eq:Mnu-mutau-separation}
\end{eqnarray}
with
\begin{eqnarray}
&&
M_{sym}  = \left( \begin{array}{*{20}c}
   M_{ee} & M^{(+)}_{e\mu } & - \sigma M^{(+)}_{e\mu }  \\
   M^{(+)}_{e\mu } & M^{(+)}_{\mu\mu } & M_{\mu \tau }   \\
    - \sigma M^{(+)}_{e\mu } & M_{\mu \tau } & M^{(+)}_{\mu\mu }\\
\end{array} \right),
\quad
M_b  = \left( \begin{array}{*{20}c}
   0 & M^{(-)}_{e\mu } & \sigma M^{(-)}_{e\mu }  \\
   M^{(-)}_{e\mu }& M^{(-)}_{\mu\mu } & 0  \\
   \sigma M^{(-)}_{e\mu } & 0 & - M^{(-)}_{\mu\mu } \\
\end{array} \right),
\label{Eq:Mnu-mutau-separation2}
\end{eqnarray}
where the obvious relations $M_{e\mu}=M^{(+)}_{e\mu}+M^{(-)}_{e\mu}$, $M_{e\tau}=- \sigma(M^{(+)}_{e\mu}-M^{(-)}_{e\mu})$, $M_{\mu\mu}=M^{(+)}_{\mu\mu}+M^{(-)}_{\mu\mu}$ and $M_{\tau\tau}= M^{(+)}_{\mu\mu}-M^{(-)}_{\mu\mu}$ are used.  It is the $\mu$-$\tau$ symmetry, where the lagrangian for $M_{sym}$: $-{\mathcal{L}}_{mass}=\psi^TM_{sym}\psi/2$ with $\psi=(\nu_e, \nu_\mu, \nu_\tau)^T$ turns out to be invariant under the exchange of $\nu_\mu\leftrightarrow -\sigma\nu_\tau$.  From $M_{sym}$, it is easy to see that the eigenvector corresponding to $\nu_3$ is given by $(0, \sigma, 1)^T/\sqrt 2$ with the eigenvalue $M^{(+)}_{\mu\mu}+\sigma M_{\mu\tau}$, leading to $s_{13}=0$ and $c_{23}=\sigma s_{23}=1/\sqrt2$ in $U_\nu$ of Eq.(\ref{Eq:U_nu}), where the sign of $\sigma$ defines the sign of $s_{23}$. Therefore, Dirac CP violation is absent in $M_{sym}$ as was announced above.  In other words, Dirac CP violation is sensitive to how the $\mu$-$\tau$ symmetry is broken \cite{MassTextureCP3}.

If the $\mu$-$\tau$ symmetry is broken by pure imaginary flavor neutrino masses, the corresponding texture can be shown to exhibit maximal Dirac CP violation as well as maximal atmospheric neutrino mixing \cite{MassTextureCP1,mu-tau-complex}.  This texture has the following form:
\begin{eqnarray}
&& 
	\left( {\begin{array}{*{20}c}
   a & b_0 & -\sigma b_0 \\
   b_0 & d_0 & e  \\
    -\sigma b_0 & e & d_0  \\
\end{array}} \right)
+i
	\left( {\begin{array}{*{20}c}
   0 & b^\prime_0 & \sigma b^\prime_0 \\
   b^\prime_0 & d^\prime_0 & 0  \\
    \sigma b^\prime_0 & 0 & -d^\prime_0  \\
\end{array}} \right),
\label{Eq:Mnu-1}
\end{eqnarray}
where all mass parameters, $a$, $b_0$, $d_0$, $b^\prime_0$, $d^\prime_0$ and $e$, are taken to be real.  There is a more general form that depends on the phase of a complex number $z$ with $\vert z\vert=1$, which is obtained by replacing ${\rm Re}(\omega)\rightarrow (\omega + z\omega^\ast)/2$ and $i{\rm Im}(\omega)\rightarrow (\omega - z\omega^\ast)/2$, where $\omega$ represents a flavor neutrino mass in Eq.(\ref{Eq:Mnu-1}). There are useful relations to estimate $\theta_{23}$ and $\delta$:
\begin{eqnarray}
&&
\tan \theta_{23}  = \frac{{\rm Im}\left( {\rm\bf M}_{e\mu}\right)}{{\rm Im}\left( {\rm\bf M}_{e\tau}\right)},
\label{Eq:theta23}
\end{eqnarray}
and
\begin{eqnarray}
s_{23} {\rm\bf M}_{e\mu} + c_{23} {\rm\bf M}_{e\tau} = \left| {s_{23} {\rm\bf M}_{e\mu} + c_{23} {\rm\bf M}_{e\tau}} \right| e^{-i\delta },
\label{Eq:Phase-delta}
\end{eqnarray}
where ${\rm\bf M}$ is a Hermitian matrix defined by ${\rm\bf M}=M^\dagger_\nu M_\nu$ and ${\rm\bf M}_{ij}$ stands for the $ij$ element of ${\rm\bf M}$ \cite{GeneralCP}.  Applying these constraints to Eq.(\ref{Eq:Mnu-1}) yields $\tan \theta_{23} = \sigma$ and $\delta = \pm\pi/2$.

To depart from the texture giving maximal Dirac CP violation and maximal atmospheric neutrino mixing is likely to give a deviation of the PMNS unitary matrix from its standard parameterization of $U^{PDG}_{PMNS}$.  For a given $M_\nu$, the PMNS unitary matrix is completely determined by three eigenvectors associated with $M_\nu$, more precisely, with the Hermitian matrix of ${\rm\bf M}$.  Namely, the form of the PMNS unitary matrix cannot be {\it a priori} assumed, but determined from $M_\nu$ \cite{Koide}.  Therefore, the phases present in three eigenvectors are not necessarily coincident with those in three columns of $U^{PDG}_{PMNS}$.  In general, there are three phases including $\delta$ associated with three rotations (except for $K$).  Of course, two of them can be rotated away by appropriate redefinition of phases of the neutrinos, which includes a redefinition of flavor neutrinos.  However, this redefinition yields a modification of $M_\nu$ so that the modified form of $M_\nu$ can be diagonalized by $U^{PDG}_{PMNS}$.  It is this modified $M_\nu$ whose three eigenvectors match with $U^{PDG}_{PMNS}$.  To obtain the modified $M_\nu$, we have to know the amount of required rotations, which add phases to relevant elements of $M_\nu$ and are precisely determined once the original $M_\nu$ is given. Therefore, if $M_\nu$ can be diagonalized by $U^{PDG}_{PMNS}$ without any rotations, its phase structure is not arbitrary but restricted.  In other words, if we impose additional constraints on the flavor neutrino masses, a given $M_\nu$ can be diagonalized by $U^{PDG}_{PMNS}$ without any rotations.\footnote{It is always true that, for a given unitary matrix $U$, any hermitian matrix $H$ can be diagonalized by $U$ if we impose artificial relations on some of elements of $H$. The correct solution is well known:  Find eigenvectors for $H$ then construct $U$.  Instead, if the most general form of $U$ is used, we can also reach the correct solution, which tells us that some of the phases vanish.}  The appearance of such extra constraints implies that the three eigenvectors associated with $M_\nu$ contain other phases than $\delta$.  It turns out that the required constraints are dictated by a set of equations for $M_\nu$ which are shown in the Appendix \ref{sec:Appendix}.

In this article, we examine a complex neutrino mass texture which can be diagonalized by $U^{PDG}_{PMNS}$ without requiring any additional constraints among the flavor neutrino masses and present four typical mass textures to be consistent with the observation of neutrino oscillations.  In Sec.\ref{sec:2}, the outline of our derivation of texture is described.  Our texture is found to provide the maximal atmospheric neutrino mixing and shows the Dirac CP violation for any value of $\delta$.  Section \ref{sec:3} deals with four textures that yield the normal and inverted mass hierarchies to realize $\Delta m^2_{atm}\gg\Delta m^2_\odot$.  We present a new texture for quasi degenerate neutrinos with their masses of ${\mathcal{O}}(\sqrt{\Delta m^2_{atm}})$, which can exhibit either $\vert m_1 \vert < \vert m_2 \vert < \vert m_3 \vert$ or $\vert m_3 \vert < \vert m_1 \vert < \vert m_2 \vert$ depending on size of a flavor neutrino mass proportional to $e^{-2i\delta}$. The final section is devoted to summary and discussions.  

\section{\label{sec:2}Texture with CP Violating Dirac Phase}
To find the appropriate texture describing CP violations for any value of $\delta$, we are basing our discussion on the two properties of complex flavor neutrino masses discussed in the previous section, which indicate the usefulness of the separation of $M_\nu$ into the $\mu$-$\tau$ symmetric and symmetry breaking part and the unique form of Eq.(\ref{Eq:Mnu-1}).  This form serves as a reference point, where the $\mu$-$\tau$ symmetry breaking part is given by pure imaginary flavor neutrino masses.  It is, thus, reasonable to presume that, for $\delta =\pm \pi/2$, the real part of $M_\nu$, which is $\mu$-$\tau$ symmetric, consists of $1$ and $e^{-2i\delta}$ and the pure imaginary part, which serves as a $\mu$-$\tau$ symmetry breaking term, consists of $e^{-i\delta}$.  These two properties suggest us to employ the following texture:
\begin{eqnarray}
&& 
M_\nu = \left( {\begin{array}{*{20}c}
   {a_0 } & {b_0 } & { - \sigma b_0 }  \\
   {b_0 } & {d_0  + d_1 e^{ - 2i\delta } } & {\sigma \left( { - d_0  + d_1 e^{ - 2i\delta } } \right)}  \\
   { - \sigma b_0 } & {\sigma \left( { - d_0  + d_1 e^{ - 2i\delta } } \right)} & {d_0  + d_1 e^{ - 2i\delta } }  \\
\end{array}} \right)
+e^{ - i\delta } 
\left( {\begin{array}{*{20}c}
   0 & {b^\prime_0 } & {\sigma b^\prime_0 }  \\
   {b^\prime_0 } & {d^\prime_0 } & 0  \\
   {\sigma b^\prime_0 } & 0 & { - d^\prime_0 }  \\
\end{array}} \right),
\label{Eq:Mnu-2}
\end{eqnarray}
where $d_1$ is an additional real mass parameter.  The suppression of $b^\prime_0$ and $d^\prime_0$ compared to the scale of $M_\nu$ is presumably of ${\mathcal{O}}(\sqrt{\Delta m^2_{atm}})$ and induces a tiny nonvanishing $\sin\theta_{13}$.  

In Eq.(\ref{Eq:Mnu-2}), the phase in the $\mu$-$\tau$ symmetry breaking sector is common to $b^\prime_0$ and $d^\prime_0$.  The appearance of the common phase may be traced back to a single $\mu$-$\tau$ symmetry breaking source in underlying dynamics such as the seesaw mechanism \cite{Seesaw}.  The underlying dynamics must not supply phases for the charged leptons, whose interactions certainly induce $\mu$-$\tau$ symmetry breaking terms for neutrinos.  This scenario is easily  realized if this single phase is given by $\mu$-$\tau$ symmetry breaking contributions to right-handed neutrino masses \cite{mu-tau2}.  If this is the case, the $e^{-i\delta}$-term appears as a common factor to specify the $\mu$-$\tau$ symmetry breaking.  Another reason to assume one phase is that if the phase varies with $b^\prime_0$ and $d^\prime_0$, $U^{PDG}_{PMNS}$ cannot be derived from $M_\nu$ of Eq.(\ref{Eq:Mnu-2}) unless we demand extra constraints on the flavor neutrino masses.  The same phase $\delta$ of the terms proportional to $e^{-2i\delta}$ also controls the phase in the $\mu$-$\tau$ symmetric part, and these terms are also required because of this reason.  However, the physical reason for the appearance of this phase is unclear at the moment and we will show in the next section textures for the inverted mass hierarchy that do not include these terms. In the Appendix \ref{sec:Appendix}, we give another derivation of Eq.(\ref{Eq:Mnu-2}) by using a solution to equations that allow us to introduce no artificial constraints.

It is remarkable that Eq.(\ref{Eq:Phase-delta}) is identically satisfied so that it consistently describes Dirac CP violation with $\delta$ embedded in this texture.  The texture Eq.(\ref{Eq:Mnu-2}) predicts the maximal atmospheric neutrino mixing because $\tan\theta_{23}=\sigma$ is obtained from Eq.(\ref{Eq:theta23}).  The convenient way to see this is to rewrite Eq.(\ref{Eq:theta23}) referring to the $\mu$-$\tau$ symmetry, and it can be readily found that
\begin{eqnarray}
&&
\tan\theta_{23}=\sigma \frac{{\rm Im} ( {\rm\bf M}^{(-)}_{e\mu} )+{\rm Im} ( {\rm\bf M}^{(+)}_{e\mu} )}{{\rm Im} ( {\rm\bf M}^{(-)}_{e\mu} )-{\rm Im} ( {\rm\bf M}^{(+)}_{e\mu} )},
\label{Eq:NonMaximalAtm}
\end{eqnarray}
where
\begin{eqnarray}
&&
{\rm\bf M}^{(+)}_{e\mu}=
M_{ee}^\ast   M_{e\mu }^{(+)}  + M_{e\mu }^{(+) \ast } ( M_{\mu \mu }^{(+)}  - \sigma M_{\mu \tau })  + M_{e\mu }^{(-) \ast } M_{\mu \mu }^{(-)},
\nonumber\\
&&
{\rm\bf M}^{(-)}_{e\mu}=
M_{ee}^\ast   M_{e\mu }^{(-)}  + M_{e\mu }^{(-) \ast } ( M_{\mu \mu }^{(+)}  + \sigma M_{\mu \tau })  + M_{e\mu }^{(+) \ast } M_{\mu \mu }^{(-)}.
\label{Eq:DeviationMaximalAtm}
\end{eqnarray}
Our texture Eq.(\ref{Eq:Mnu-2}) has real flavor neutrino masses of $M_{ee}$, $M^{(+)}_{e\mu}$ and $M^{(+)}_{\mu\mu}-\sigma M_{\mu\tau}$ while $M^{(-)}_{e\mu,\mu\mu}$ have the common phase.  We then have ${\rm Im} ( {\rm\bf M}^{(+)}_{e\mu})=0$, which gives $\tan\theta_{23}=\sigma$ and the maximal atmospheric neutrino mixing shows up.

From Eq.(\ref{Eq:theta13-single}) in Appendix \ref{sec:Appendix}, the mixing angles $\theta_{12,13}$ are expressed as
\begin{eqnarray}
&&
\tan 2\theta _{12}  = 2\sqrt 2\frac{1}{c_{13}} \frac{M_{e\mu }^{( + )} }{\lambda _2  - \lambda _1},
\quad
\tan 2\theta _{13}  = 2\sqrt 2 \frac{\sigma M_{e\mu }^{( - )} e^{i\delta } }{\lambda _3 e^{ 2i\delta }  - a_0 },
\label{Eq:Theta12-13}
\end{eqnarray}
which reproduce Eq.(\ref{Eq:Results}).  The phase completely disappears to yield real values of $\tan 2\theta _{12}$ and $\tan 2\theta _{13}$.  This is because our texture gives the following mass parameters:
\begin{eqnarray}
&&
M_{e\mu }^{( + )}= b_0,
\quad
M_{e\mu }^{( - )} = \sigma e^{ - i\delta }b^\prime_0,
\nonumber\\
&& 
\lambda _1  = \frac{c_{13}^2 a_0 - 2s_{13}^2 d_1}{c_{13}^2  - s_{13}^2},
\quad
\lambda _2  = 2d_0,
\quad
\lambda _3  = 2 e^{ - 2i\delta }d_1.
\label{Eq:MassParameters}
\end{eqnarray}
We then find that $\lambda_{1,2}$, $\lambda _3 e^{2i\delta }$, $M_{e\mu }^{( + )}$, and $M_{e\mu }^{( - )} e^{i\delta }$ become real so that $\tan 2\theta _{12}$ and $\tan 2\theta _{13}$ become real.  If the phase differs in $b^\prime_0$ and $d^\prime_0$, $\lambda_1$ ceases to be real and the phase of $M_{e\mu }^{( - )}$ is not $\delta$.  The imaginary parts of $\tan 2\theta _{12}$ and $\tan 2\theta _{13}$  should vanish, and this gives extra constraints among the flavor neutrino masses.\footnote{These constraints turn out to be fictitious if additional phases are introduced, and these phases are the correct ingredients for the $PMNS$ unitary matrix in this case \cite{New}.}

We can examine Majorana phases predicted in Eq.(\ref{Eq:neutrino-masses}).  Since $\lambda_{1,2}$ and $X (= \sqrt{2}M_{e\mu }^{( + )}/c_{13})$ turn out to be real, the Majorana phases of $\beta_{1,2}$ vanish in Eq.(\ref{Eq:neutrino-masses}).  On the other hand, we find that
\begin{eqnarray}
&&
m_3 e^{ - 2i\beta _3 }  = \frac{{2c_{13}^2 d_1  - s_{13}^2 a_0}}{{c_{13}^2  - s_{13}^2 }}e^{ - 2i\delta },
\label{Eq:ThirdMass}
\end{eqnarray}
which yields $\beta _3 = \delta$.  The Majorana phase associated with the third massive neutrino is identical to the CP violating Dirac phase.

\section{\label{sec:3}Neutrino Mass Hierarchies}
In this section, we discuss how the observed properties of neutrino oscillations such as $\sin^2\theta_{13}\ll 1$ and $\Delta m^2_{atm}\gg \Delta m^2_\odot$ are explained in our proposed texture.  For $\sin^2\theta_{13}\ll 1$, we know that any effect from the $\mu$-$\tau$ symmetry breaking to be denoted by $\varepsilon$ should be tiny and characterizes $b^\prime_0$ and $d^\prime_0$.  For $\Delta m^2_{atm}\gg \Delta m^2_\odot$, we present explicit forms of textures in the normal and inverted mass hierarchies because the degenerate neutrino mass pattern cannot provide $\Delta m^2_{atm}\gg \Delta m^2_\odot$.  Three Majorana phases are fixed to be $\beta_{1,2}=0$ and $\beta_3=\delta$.  Neutrino masses are then  calculated from
\begin{eqnarray}
&&
m_1  \approx \frac{{a_0  + 2d_0  + \left( {a_0  - 2d_1 } \right)t_{13}^2 }}{2} - \frac{{\sqrt 2 b_0 }}{{\sin 2\theta _{12} }}\left(1+\frac{1}{2}t^2_{13}\right),
\nonumber\\
&&
m_2  \approx \frac{{a_0  + 2d_0  + \left( {a_0  - 2d_1 } \right)t_{13}^2 }}{2} + \frac{{\sqrt 2 b_0 }}{{\sin 2\theta _{12} }}\left(1+\frac{1}{2}t^2_{13}\right),
\nonumber\\
&&
m_3  \approx {2d_1  - \left( {a_0 - 2d_1} \right)t_{13}^2 },
\label{Eq:CalculatedMasses}
\end{eqnarray}
where the terms of ${\mathcal{O}}(\sin^2\theta_{13})$ are properly taken into account. The effect of these terms is not significant in the normal mass hierarchy but comparable to the effect from the other terms in the inverted mass hierarchy.  Furthermore, it will be shown that their effect can provide $\Delta m^2_\odot/\Delta m^2_{atm} \sim \sin^2\theta_{13}$ \cite{Theta31AndMass} in a new type of texture for quasi degenerate neutrinos with masses of ${\mathcal{O}}(\sqrt{\Delta m^2_{atm}})$.  We denote parameters used in $M_{ee}$ to be $p$, in $M_{e\mu}$ to be $q$, in $M_{\mu\mu}$ to be $r$.

\subsection{\label{Eq:subsec:3-1}Normal Mass Hierarchy}
Our texture can be parameterized by $b_0  = \eta d_1$, $b^\prime_0  = \varepsilon d_1$, $a_0  = p\eta d_1$, $d_0  = r\eta d_1$ and $d^\prime_0  = x\varepsilon d_1$, which results in
\begin{eqnarray}
&&
M_\nu   = d_1 \left( {\begin{array}{*{20}c}
   {p\eta } & \eta + \varepsilon e^{ - i\delta } &  - \sigma  \left( \eta - \varepsilon e^{ - i\delta } \right)  \\
   \eta + \varepsilon e^{ - i\delta } &  r\eta + x\varepsilon e^{ - i\delta } + e^{ - 2i\delta }  & -\sigma \left( r\eta  - e^{ - 2i\delta }  \right)  \\
    - \sigma  \left( {\eta - \varepsilon e^{ - i\delta } } \right) & -\sigma \left( r\eta  - e^{ - 2i\delta }  \right) & r\eta - x\varepsilon e^{ - i\delta } + e^{ - 2i\delta } \\
\end{array}} \right),
\label{Eq:Normal}
\end{eqnarray}
where $\eta$ satisfies $\vert\eta\vert\ll 1$ and $x$ is determined so as to satisfy Eqs.(\ref{Eq:Results}) and (\ref{Eq:Results2}) for $\theta_{13}$.  This texture has often been discussed in the literatures \cite{KnownNormal}.  The mixing angles and masses are then calculated to be:
\begin{eqnarray}
&&
\tan 2\theta _{12}  \approx \frac{2\sqrt 2\eta }{\left(2r - p\right)\eta+2t^2_{13}},
\quad
\tan 2\theta _{13}  \approx \sqrt 2 \sigma \varepsilon,
\label{Eq:NormalMixing}\\
&&
m_1 \approx \left(\frac{\left(p + 2r\right)\eta-2t^2_{13}}{2} - \frac{\sqrt 2 \eta }{\sin 2\theta _{12}}\right)d_1, 
\quad
m_2 \approx \left(\frac{\left(p + 2r\right)\eta-2t^2_{13}}{2} + \frac{\sqrt 2 \eta}{\sin 2\theta _{12}} \right)d_1,
\nonumber\\
&&
m_3 \approx 2d_1,
\label{Eq:NormalMasses}\\
&&
\Delta m_ \odot ^2  \approx  \frac{2\sqrt 2\left(\left(p + 2r\right)\eta-2t^2_{13}\right)}{\sin 2\theta _{12} }\eta d_1^2,
\quad
\Delta m_{atm}^2  \approx 4d_1^2.
\label{Eq:NormalMassesSquared}
\end{eqnarray}
Since we numerically know that $t^2_{13}\mapleq \Delta m^2_\odot/\Delta m^2_{atm}$, $t^2_{13}$ gives a minor contribution to $\Delta m_ \odot ^2 $.  Then, we obtain
\begin{eqnarray}
&&
\tan 2\theta _{13}  = {\mathcal{O}}\left( \varepsilon \right),
\quad
\eta  = {\mathcal{O}}\left( \sqrt {\Delta m_ \odot ^2/\Delta m_{atm}^2 }  \right).
\label{Eq:NormalAngle13}
\end{eqnarray}
Form these results, we find that either $p$ or $r$ can be set to zero.  In this texture, it is well known that $\vert M_{ij}\vert\gg\vert M_{ei}\vert\gg\vert M_{ee}\vert$ ($i,j=\mu,\tau$) can be ascribed to the tiny violation of the electron number conservation in leptonic interactions \cite{eNumber}.  In this case, $p\sim \eta$ is anticipated and can be neglected.

There is a new texture with $m_1\sim - m_2$, which yields $\Delta m^2_\odot\sim \sin^2\theta_{13}\Delta m^2_{atm}$.\footnote{This case corresponds to quasi degenerate mass pattern with $m^2_1,m^2_2,m^2_3\gg \Delta m^2_\odot$ but not to the one with $m^2_1,m^2_2,m^2_3\gg \Delta m^2_{atm}$.  We list this pattern as the normal mass hierarchy.}  The texture is characterized by $a_0  =  - 2d_0$, $b_0  = qd_0$, $b^\prime_0  = \varepsilon d_0$, $d_1  =  - rd_0$ and $d^\prime_0  = x\varepsilon d_0$, which gives
\begin{eqnarray}
&&
M_\nu   =
d_0 \left( {\begin{array}{*{20}c}
   - 2 & q + \varepsilon e^{ - i\delta } &  - \sigma \left( q - \varepsilon e^{ - i\delta }  \right)  \\
   q + \varepsilon e^{ - i\delta }  & {1 + x\varepsilon e^{ - i\delta }  - re^{ - 2i\delta } } & -\sigma \left( 1 + re^{ - 2i\delta } \right)  \\
    - \sigma \left( q - \varepsilon e^{ - i\delta }  \right) & -\sigma \left(  1 + re^{ - 2i\delta }  \right) & {1 - x\varepsilon e^{ - i\delta }  - re^{ - 2i\delta } }  \\
\end{array}} \right).
\label{Eq:Normal2}
\end{eqnarray}
The mixing angles and masses are then calculated to be:
\begin{eqnarray}
&&
\tan 2\theta _{12}  \approx \frac{q}{{\sqrt 2 }},
\quad
\tan 2\theta _{13}  \approx -\frac{{\sqrt 2 }}{r-1}\sigma \varepsilon,
\label{Eq:Normal2Mixing}\\
&&
m_1 \approx - \left( { \frac{2}{{\cos 2\theta _{12} }} - \left( {r - 1} \right)t_{13}^2 } \right)d_0, 
\quad
m_2 \approx \left( {\frac{2}{{\cos 2\theta _{12} }}  + \left( {r - 1} \right)t_{13}^2 } \right)d_0, 
\nonumber\\
&&
m_3 \approx -2rd_0,
\label{Eq:Normal2Masses}\\
&&
\Delta m_ \odot ^2  \approx \frac{8\left(r - 1 \right) }{\cos 2\theta _{12} }t_{13}^2d_0^2,
\quad
\Delta m_{atm}^2  \approx 4\bigg| 
 r^2  - \frac{1}{\cos ^2 2\theta _{12} } \bigg|d_0^2,
\label{Eq:Normal2MassesSquared}
\end{eqnarray}
where $r>1$ for $\Delta m_ \odot ^2>0$ and $q$ appearing in $m_{1,2}$ is replaced by $\tan 2\theta_{12}$ in Eq.(\ref{Eq:Normal2Mixing}) . We obtain that
\begin{eqnarray}
&&
\tan 2\theta _{13}  
= {\mathcal{O}}\left( \varepsilon \right)
= {\mathcal{O}}\left( \sqrt {\Delta m_ \odot ^2/\Delta m_{atm}^2 }  \right),
\label{Eq:Normal2Angle13}
\end{eqnarray}
which indicates that
\begin{eqnarray}
&&
\Delta m_ \odot ^2  \sim \sin^2\theta_{13}\Delta m_{atm}^2.
\label{Eq:Normal2MassesSquared13}
\end{eqnarray}
This prediction lies in the right range of $\Delta m_ \odot ^2$ for $\sin^2\theta_{13} ={\mathcal{O}}(10^{-2})$ since $\sin^2\theta_{13} \mapleq 0.03$.

\begin{figure}[!htbp]
\begin{flushleft}
\includegraphics*[20mm,186mm][300mm,265mm]{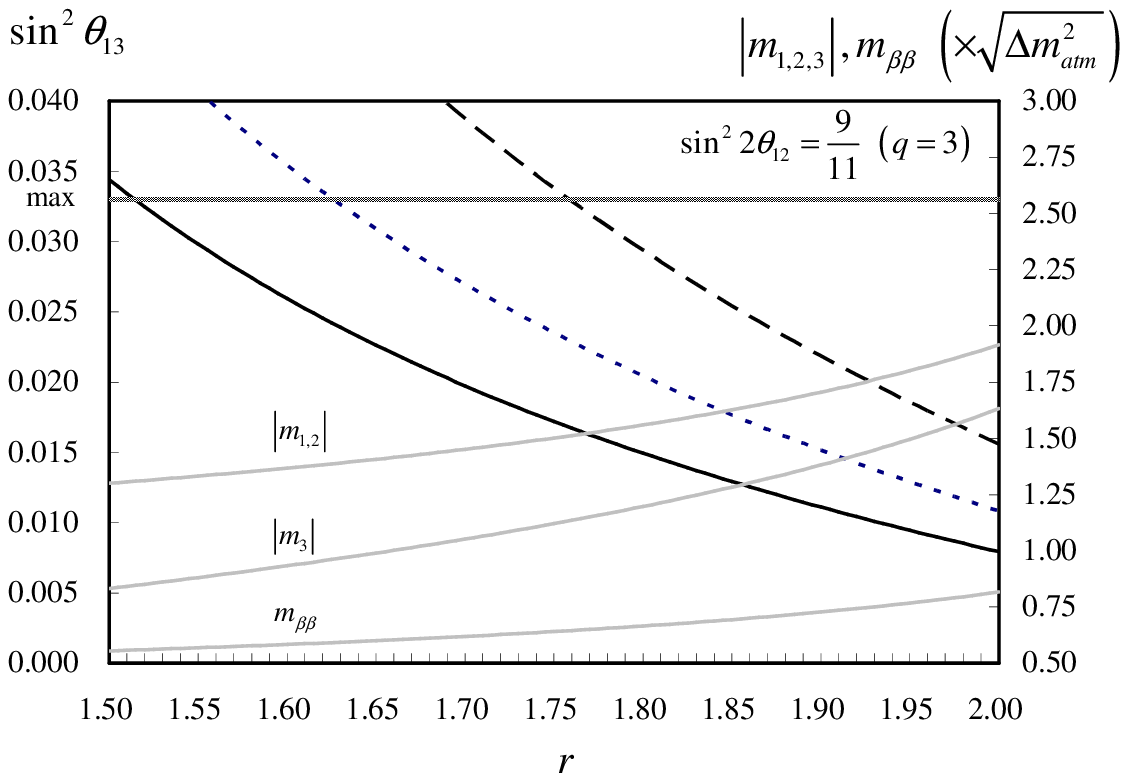}
\end{flushleft}
\caption{The predictions of $\sin^2\theta_{13}$ (black curves) and $\vert m_{1,2,3}\vert,m_{\beta\beta}$ in unit of $\sqrt{\Delta m^2_{atm}}$ (grey curves) as functions of $r$ ($\leq 2$) and $\Delta m^2_\odot/\Delta m^2_{atm}(=R)$ for $\vert m_3\vert < \vert m_{1,2}\vert$, where the solid, dotted and dashed curves, respectively, correspond to the lower bound of $R$, the center value of $R$ and the upper bound of $R$, and the upper, middle and lower grey curves, respectively, stand for the cases of $\vert m_1\vert\approx \vert m_2\vert$, $\vert m_3\vert$ and $m_{\beta\beta}$.  The upper bound on $\sin^2\theta_{13}$ is indicated by the horizontal line marked as ``max". The parameter $q$ is fixed to be 3 giving $\sin^2 2\theta_{12}$=$9/11$.}
\label{Fig:normal-2}
\end{figure}

\begin{figure}[!htbp]
\begin{flushleft}
\includegraphics*[20mm,186mm][300mm,265mm]{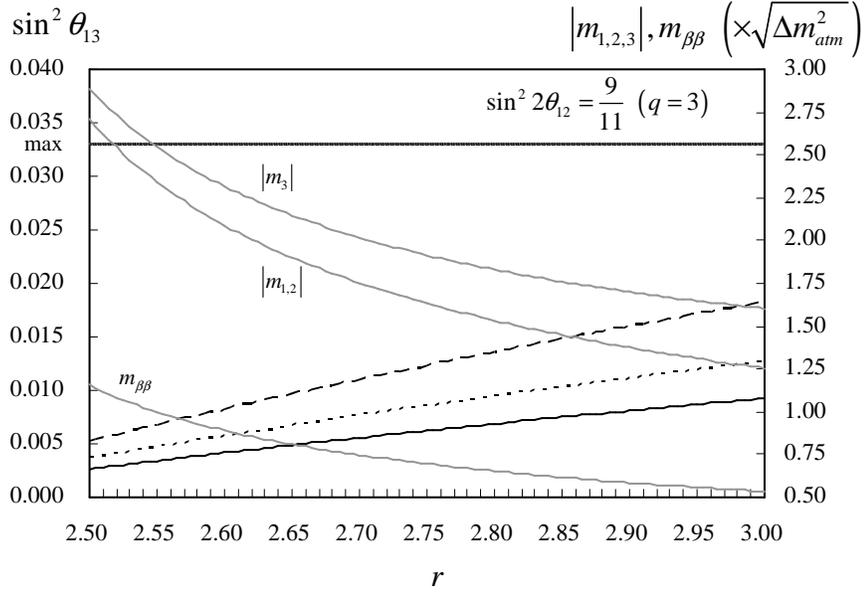}
\end{flushleft}
\caption{The same as in FIG.\ref{Fig:normal-2} but for $r\geq 2.5$ and $\vert m_3\vert > \vert m_{1,2}\vert$, where the upper and middle grey curves, respectively, stand for the cases of $\vert m_3\vert$ and $\vert m_1\vert\approx \vert m_2\vert$.}
\label{Fig:normal-2-a}
\end{figure}

Illustrated in FIG.\ref{Fig:normal-2} for $1.5\leq r \leq 2$ and in FIG.\ref{Fig:normal-2-a} for $2\leq r \leq 2.5$ are our predictions of $\sin^2\theta_{13}$ and masses including the effective neutrino mass $m_{\beta\beta}$ \cite{TheoryMass-ee} used in the detection of the absolute neutrino mass \cite{AbsoluteMass}, which is equal to $\vert M_{ee}\vert$.  The parameter $q$ is fixed to be $q=3$ giving $\sin^22\theta_{12}=9/11$.  Although this texture seems to describe the normal mass hierarchy, these two ranges of $r$ give different mass orderings: $\vert m_3 \vert < \vert m_1 \vert < \vert m_2 \vert$ for $1.5\leq r \leq 2$ and $\vert m_1 \vert < \vert m_2 \vert < \vert m_3 \vert$ for $2.5\leq r \leq 3$.  The turning point is $r\approx 1/\cos 2\theta_{12}$, which is about 2.35.  These figures indicate the following features of the texture:
\begin{itemize}
\item The prediction of $\sin^2\theta_{13}$ is obtained for the experimentally allowed values of $\Delta m^2_\odot/\Delta m^2_{atm}$ and shows that $0.010\mapleq\sin^2\theta_{13}\mapleq 0.032$ for $\vert m_3\vert < \vert m_{1,2}\vert$ and $0.005\mapleq\sin^2\theta_{13}\mapleq 0.015$ for $\vert m_3\vert > \vert m_{1,2}\vert$.
\item The degeneracy of $\vert m_{1,2,3}\vert$ shows up. Although the two types of the mass ordering satisfy either $\vert m_1 \vert < \vert m_2 \vert < \vert m_3 \vert$ or $\vert m_3 \vert < \vert m_1 \vert < \vert m_2 \vert$, the mass ordering does not correspond to either the normal or inverted mass hierarchies in a strict sense because quasi degenerate neutrinos have masses of ${\mathcal{O}}(\sqrt{\Delta m^2_{atm}})$.
\item  The effective neutrino mass $m_{\beta\beta}$ is also estimated to be: $m_{\beta\beta}={\mathcal{O}}(\sqrt{\Delta m^2_{atm}})$ even for the normal mass hierarchy.  It should be noted that most of existing textures predict $m_{\beta\beta}\ll {\mathcal{O}}(\sqrt{\Delta m^2_{atm}})$ for the normal mass hierarchy because $M_{ee}$ is suppressed as in Eq.(\ref{Eq:Normal}) \cite{TheoryMass-ee}.  In this texture, we predict that 
\begin{eqnarray}
&&
m_{\beta\beta}\sim {\rm a~few}~\times 10^{-2}~({\rm eV}),
\label{Eq:Normal2Masses-ee}
\end{eqnarray}
for $\sqrt{\Delta m^2_{atm}}\sim 5 \times 10^{-2}$ eV, even for the case $\vert m_1 \vert < \vert m_2 \vert < \vert m_3 \vert$.
\end{itemize}
These features are specific to this texture, including both $\vert m_1 \vert < \vert m_2 \vert < \vert m_3 \vert$ and $\vert m_3 \vert < \vert m_1 \vert < \vert m_2 \vert$ depending on the size of $r$. It should be stressed that $re^{-2i\delta}$ is a new term for CP-violation.

\subsection{\label{Eq:subsec:3-2}Inverted Mass Hierarchy}
There are two different textures depending on the relative sign of $m_1$ and $m_2$: $m_1 \sim m_2$ or $m_1 \sim -m_2$ \cite{PlusMinusNu}.  Our texture with $m_1 \sim m_2$ can be parameterized by $a_0  = 2d_0 ( {1 - p\eta } )$, $b_0  = \eta d_0$, $b^\prime_0  = \varepsilon d_0$, $d_1  = 0$ and $d^\prime_0  = x\varepsilon d_0$, which gives
\begin{eqnarray}
&&
M_\nu   =d_0 \left( {\begin{array}{*{20}c}
   {2\left( {1 - p\eta } \right)} & \eta + \varepsilon e^{ - i\delta } &  - \sigma \left( {\eta - \varepsilon e^{ - i\delta } } \right)  \\
   \eta  + \varepsilon e^{ - i\delta } & {1 + x\varepsilon e^{ - i\delta } } & { - \sigma }  \\
   { - \sigma \left( {\eta  - \varepsilon e^{ - i\delta } } \right)} & { - \sigma } & {1 - x\varepsilon e^{ - i\delta } }  \\
\end{array}} \right).
\label{Eq:Inverted}
\end{eqnarray}
The mixing angles and masses are then calculated to be:
\begin{eqnarray}
&&
\tan 2\theta _{12}  \approx \frac{\sqrt 2\eta}{p\eta-t^2_{13}},
\quad
\tan 2\theta _{13}  \approx - \sqrt 2 \sigma \varepsilon,
\label{Eq:InvertedMixing}\\
&&
m_1 \approx \left( {2 - \sqrt 2 \frac{{\cos 2\theta _{12}  + 1}}{{\sin 2\theta _{12} }}\eta } \right)d_0, 
\quad
m_2 \approx \left( {2 - \sqrt 2 \frac{{\cos 2\theta _{12}  - 1}}{{\sin 2\theta _{12} }}\eta } \right)d_0, 
\nonumber\\
&&
m_3 \approx  - 2t^2_{13} d_0,
\label{Eq:InvertedMasses}\\
&&
\Delta m_ \odot ^2  \approx \frac{{8\sqrt 2 }}{{\sin 2\theta _{12} }}\eta d_0^2,\quad
\Delta m_{atm}^2  \approx 4d_0^2,
\label{Eq:InvertedMassesSquared}
\end{eqnarray}
where the term $p\eta-t^2_{13}$ appearing in $m_{1,2}$ has been replaced by $\tan 2\theta_{12}$. Then, we obtain
\begin{eqnarray}
&&
\tan 2\theta _{13}  = {\mathcal{O}}\left( \varepsilon\right),
\quad
\eta  = {\mathcal{O}}\left( \Delta m_ \odot ^2 /\Delta m_{atm}^2 \right).
\label{Eq:InvertedAngle13}
\end{eqnarray}
This estimation of $\eta$ shows that the $t^2_{13}$-term in $\tan 2\theta_{12}$ may be comparable to $\eta$ and cannot be neglected. 

Another texture, which shows $m_1\sim -m_2$, has $a_0  =  - 2d_0 ( {1 - \eta })$. $b_0  = qd_0$, $b^\prime_0  = \varepsilon d_0 $, $d_1  = 0$ and $d^\prime_1  = x\varepsilon d_0$ and is given by
\begin{eqnarray}
&&
M_\nu   =
d_0 \left( {\begin{array}{*{20}c}
    - 2\left( {1 - \eta } \right) & q + \varepsilon e^{ - i\delta }  & - \sigma \left( q - \varepsilon e^{ - i\delta }  \right)  \\
   q + \varepsilon e^{ - i\delta } & 1 + x\varepsilon e^{ - i\delta }  & { - \sigma }  \\
    - \sigma \left( q - \varepsilon e^{ - i\delta }  \right) & { - \sigma } & 1 - x\varepsilon e^{ - i\delta }   \\
\end{array}} \right).
\label{Eq:Inverted2}
\end{eqnarray}
The mixing angles and masses are then calculated to be:
\begin{eqnarray}
&&
\tan 2\theta _{12}  \approx \frac{q}{\sqrt 2 },
\quad
\tan 2\theta _{13}  \approx \sqrt 2 \sigma \varepsilon,
\label{Eq:Inverted2Mixing}\\
&&
m_1 \approx  - \left( \frac{2}{\cos 2\theta _{12} } - \eta+t^2_{13} \right)d_0 ,
\quad
m_2  \approx \left( \frac{2}{\cos 2\theta _{12} } +\eta-t^2_{13} \right)d_0 , 
\nonumber\\
&&
m_3  \approx 2t^2_{13} d_0,
\label{Eq:Inverted2Masses}\\
&&
\Delta m_ \odot ^2  \approx \frac{8\left(\eta-t^2_{13}\right)}{\cos 2\theta _{12} } d_0^2,
\quad
\Delta m_{atm}^2  \approx \frac{4 }{\cos ^2 2\theta _{12} }d_0^2,
\label{Eq:Inverted2MassesSquared}
\end{eqnarray}
where $q$ in $m_{1,2}$ is replaced by $\tan 2\theta_{12}$ because of Eq.(\ref{Eq:Inverted2Mixing}).  As in the previous cases, we obtain
\begin{eqnarray}
&&
\tan 2\theta _{13}  = {\mathcal{O}}\left( {\varepsilon } \right).
\label{Eq:Inverted2Angle13}
\end{eqnarray}
Because $t^2_{13}\mapleq \Delta m^2_\odot/\Delta m^2_{atm}$, we cannot neglect the terms proportional to $t^2_{13}$ in Eq.(\ref{Eq:Inverted2MassesSquared}) for $\Delta m_ \odot ^2$ as long as $t^2_{13}\sim 10^{-2}$.  We, therefore, expect that
\begin{eqnarray}
&&
\eta  = {\mathcal{O}}\left( \Delta m_ \odot ^2 /\Delta m_{atm}^2 \right),
\label{Eq:Inverted2Eta}
\end{eqnarray}
with $\eta>t^2_{13}$ to satisfy $\Delta m_ \odot ^2 > 0$.

In these four textures, the feature of $\tan 2\theta_{13}={\mathcal{O}}({\varepsilon})$ appears as a direct consequence of Eq.(\ref{Eq:theta13-single}) giving $\tan 2\theta_{13} \propto M^{(-)}_{e\mu}$ because of $c_{23}=\sigma s_{23}=1/\sqrt 2$.  The similar feature also  appears in the predictions of $\tan 2\theta_{12} \propto M^{(+)}_{e\mu}$ from the textures Eqs.(\ref{Eq:Normal}) and (\ref{Eq:Inverted}), where another small parameter $\eta$ apparantly gives $\tan 2\theta_{12}={\mathcal{O}}({\eta})$.  However, the contribution from ${\mathcal{O}}(\eta)$ in the numerator is cancelled by the same factor in the denominator as in Eq.(\ref{Eq:NormalMixing}) to give $\tan 2\theta_{12}$ of order one.  If we demand that $\eta\sim\varepsilon$, we can find $\Delta m^2_\odot\sim t^2_{13}\Delta m_{atm}^2$ in the normal mass hierarchy of Eq.(\ref{Eq:Normal}) or $\Delta m^2_\odot\sim t_{13}\Delta m_{atm}^2$ for the inverted mass hierarchy of Eqs.(\ref{Eq:Inverted}) and (\ref{Eq:Inverted2}) \cite{Theta31AndMass}.  However, there is no {\it a priori} theoretical reason to suppose $\eta \sim \varepsilon$, although the present experimental data suggest it.  It should be emphasized that the generic prediction of $\Delta m^2_\odot\sim \sin^2\theta_{13}\Delta m_{atm}^2$ produced by the terms of ${\mathcal{O}}(\sin^2\theta_{13})$ is specific to the new type of the texture Eq.(\ref{Eq:Normal2}).

\section{\label{sec:4}Summary and Discussions}
We have successively demonstrated that the proposed texture of Eq.(\ref{Eq:Mnu-2}) describes the maximal atmospheric neutrino mixing with arbitrary CP violating Dirac phase $\delta$, as well as the observed properties of neutrino oscillations such as $\Delta m^2_{atm}\gg \Delta m^2_\odot$.  The general and simple formula to estimate the deviation from the maximal atmospheric neutrino mixing has been given in Eq.(\ref{Eq:NonMaximalAtm}), which utilizes the classification of $M^\dagger_\nu M_\nu$ due to the $\mu$-$\tau$ symmetry.  The source giving the deviation is the imaginary part of ${\rm\bf M}^{(+)}_{e\mu}=M_{ee}^\ast   M_{e\mu }^{(+)}  + M_{e\mu }^{(+) \ast } ( M_{\mu \mu }^{(+)}  - \sigma M_{\mu \tau })  + M_{e\mu }^{(-) \ast } M_{\mu \mu }^{(-)}$. The present texture gives ${\rm Im}({\rm\bf M}^{(+)}_{e\mu})=0$ and so $\tan\theta_{23}=\sigma$ is derived.  In addition to the CP violating Dirac phase, the Majorana phase associated with the third neutrino becomes $\delta$, and other Majorana phases vanish.  Our texture, thus, yields Majorana CP violation. To reach this texture, we have relied upon the usefulness of the $\mu$-$\tau$ symmetry, which allows us to divide a given texture into the symmetric part and the symmetric breaking part.  Since the effect of Dirac CP violation arises from the symmetric breaking part, we characterize this part to be proportional to $\varepsilon e^{-i\delta}$.  One may wonder what happens if $\varepsilon =0$ in Eq.(\ref{Eq:Mnu-2}), which still exhibits the phase $\delta$.  Since the $\mu$-$\tau$ symmetric texture gives $\sin\theta_{13}=0$, there is no Dirac CP violation phase $\delta$ in $U^{PDG}_{PMNS}$.  In terms of the flavor neutrino masses, we see that ${\rm Im}\left({\rm\bf M}_{e\mu}\right)={\rm Im}\left({\rm\bf M}_{e\tau}\right)=0$, which jeopardizes the validity of Eq.(\ref{Eq:Phase-delta}) to indicate no Dirac CP violation, and Eq.(\ref{Eq:theta23}) is replaced by $\tan\theta_{23}=-{\rm Re}\left( {\rm\bf M}_{e\tau}\right)/{\rm Re}\left( {\rm\bf M}_{e\mu}\right)$ \cite{mu-tau-complex}. The phase $\delta$ embedded in our $\mu$-$\tau$ symmetric part of the texture is transferred to the Majorana phase, which is identical to $\delta$ as can be seen from Eq.(\ref{Eq:ThirdMass}).

As stated in Sec.\ref{sec:2}, it is our main assumption that the single phase $\delta$ controls the $\mu$-$\tau$ symmetry breaking part.  The theoretical reason is to confine ourselves within the textures that precisely have $U^{PDG}_{PMNS}$ as the PMNS unitary matrix without entailing further rotations due to the redefinition of flavor neutrinos.  However, the inclusion of the additional phase $\alpha$ defined in Eq.(\ref{Eq:mu-tau-related}) as a free parameter ($\alpha\neq -\delta$) practically does not alter most of our main conclusions since the $\mu$-$\tau$ symmetry breaking causes very tiny effects.  One direct consequence is that the atmospheric neutrino mixing ceases to be maximal as expected.  In our case, it is determined by Eq.(\ref{Eq:theta23-single}), which provides $\cos 2\theta_{23} (\sim t_{13}) \sim \varepsilon$ because of the mismatch of the phase of $X$ with $\delta$, as can be seen from Eq.(\ref{Eq:theta13-single}).\footnote{In fact, we can find that $\cos 2\theta_{23}\propto [d_0(c_\delta   - c_\alpha ) + d_1( c_\delta   - c_{2\delta  - \alpha })]d^\prime_0$ \cite{New}.} 

It is recognized that
\begin{itemize}
\item the $\mu$-$\tau$ symmetric part gives $\tan 2\theta_{12} \propto M^{(+)}_{e\mu}$, and
\item the $\mu$-$\tau$ symmetry breaking part gives $\tan 2\theta_{13} \propto M^{(-)}_{e\mu}$.
\end{itemize}
To obtain this contrasted result shows the usefulness of the classification due to the $\mu$-$\tau$ symmetry.\footnote{The generic use of the $\mu$-$\tau$ symmetry reveals that the $\mu$-$\tau$ symmetric texture does not necessarily lead to $\sin\theta_{13}=0$ but to $\sin\theta_{12}=0$ depending on the mass ordering of three eigenvalues for $M_\nu$ \cite{GeneralCP}. Namely, the eigenvector corresponding to $\nu_3$ proportional to $(u, 1, -\sigma)^T$ yields $\sin\theta_{13} \propto u$ and $\sin\theta_{12}=0$, while $(0, \sigma, 1)^T$ gives $\sin\theta_{13}=0$. The detailed discussions will be presented elsewhere \cite{New}.}  Furthermore, we note that the relation $M^{(-)}_{\mu\mu }  =  - \sqrt 2 \sigma t_{13} e^{ - i\delta } M_{e\mu}^{(+)}$ as in Eq.(\ref{Eq:ExamplesConstraint}) gives a definite correlation between the $\mu$-$\tau$ symmetric part $M_{e\mu}^{(+)}$ and its symmetry breaking part $M^{(-)}_{\mu\mu }$.  This relation can be viewd as a main constraint to have the maximal atmospheric neutrino mixng. 

Four types of realization of $\Delta m^2_{atm}\gg \Delta m^2_\odot$ are explicitly given by specific textures.  Among others, we have found a new type of realization, which uses terms of ${\mathcal{O}}(\sin^2\theta_{13})$.  Namely, we have shown that
\begin{eqnarray}
&&
\Delta m^2_\odot\sim \sin^2\theta_{13}\Delta m_{atm}^2,
\label{Eq:Solar_vs_Atm}
\end{eqnarray}
which shows the right order of the observed hierarchy because $\sin^2\theta_{13} \mapleq 0.03$.  Namely, $\sin^2\theta_{13}={\mathcal{O}}(10^{-2})$ is predicted in this case. It is the only case, where $\Delta m^2_\odot/\Delta m_{atm}^2$ is induced by the effect from the terms of ${\mathcal{O}}(\sin^2\theta_{13})$.  The explicit form for $\varepsilon\rightarrow 0$ is given by
\begin{eqnarray}
&&
M_\nu   =
d_0 \left( {\begin{array}{*{20}c}
    - 2 & q &  - \sigma q  \\
   q & 1 - re^{ - 2i\delta }  & -\sigma \left( 1 + re^{ - 2i\delta }  \right)  \\
    - \sigma q & -\sigma \left(  1 + re^{ - 2i\delta }  \right) & 1 - re^{ - 2i\delta }   \\
\end{array}} \right).
\label{Eq:Normal2Summary}
\end{eqnarray}
This texture is found to have the following features:
\begin{itemize}
\item Neutrinos are quasi degenerate ones with masses of ${\mathcal{O}}(\sqrt{\Delta m^2_{atm}})$,\item Depending on the size of $r$, the mass ordering becomes either $\vert m_1 \vert < \vert m_2 \vert < \vert m_3 \vert$ or $\vert m_3 \vert < \vert m_1 \vert < \vert m_2 \vert$,
\item The effective neutrino mass $m_{\beta\beta}$ is a few $\times 10^{-2}$ eV.
\end{itemize}
It should be noted that the textures with $\delta=0$ also give the right answers to describe the observed neutrino oscillations because the phases do not contribute in the estmation of the masses and the mixing angles.  However, the atmospheric neutrino mixing is not maximal because Eq.(\ref{Eq:theta23}) is not satisfied. This is due to the lack of the constraint from the imaginary part of Eq.(\ref{Eq:theta23-single}).  The mixing angle $\theta_{23}$ is determined by Eq.(\ref{Eq:theta23-single}) with $\delta=0$ corresponding to the real part of Eq.(\ref{Eq:theta23-single}), which implies that $M_{\mu\mu}-M_{\tau\tau} \propto s_{13}$ and $\cos 2\theta_{23}\propto s_{13}$.

The complex flavor neutrino mass matrix with its elements satisfying Eq.(\ref{Eq:Solution}) is the unique form of the matrix describing the maximal atmospheric neutrino mixing that can be diagonalized by $U^{PDG}_{PMNS}$ without requiring other constraints among the flavor neutrino masses.  A typical example is the texture discussed here, where the Dirac CP violation is described by an arbitrary phase $\delta$ embedded in the mass matrix.  To find phenomenologically viable neutrino mass textures, a specific type of the $\mu$-$\tau$ symmetry breaking Eq.(\ref{Eq:Examples}) is chosen and four types of textures are shown to consistently describe the neutrino mass hierarchy.  As discussed in Sec.\ref{sec:2}, it is likely that underlying dynamics provides a single $\mu$-$\tau$ symmetry breaking source with the phase $\delta$, which can induce the terms proportional to $e^{-i\delta}$.  In this case, Eqs.(\ref{Eq:Inverted}) and (\ref{Eq:Inverted2}) for the inverted mass hierarchy match this scenario because the additional terms proportional to $e^{-2i\delta}$ are absent.  One can construct other types of mass matrices than Eq.(\ref{Eq:Mnu-2}) if one chooses other types of the $\mu$-$\tau$ symmetry breaking because the CP violation is controlled by the part of the textures which is not $\mu$-$\tau$ symmetric.  Other useful patterns of the $\mu$-$\tau$ symmetry breaking part will be discussed elsewhere \cite{New}.

\vspace{3mm}
\noindent
\centerline{\small \bf ACKNOWLEGMENTS}

The authors are grateful to T. Kitabayashi for enlightening discussions and to W. Bentz for reading the manuscript and for useful comments.  The work of M.Y. is supported by the Grants-in-Aid for Scientific Research on Priority Areas (No 13135219) from the Ministry of Education, Culture, Sports, Science, and Technology, Japan. 

\appendix
\section{\label{sec:Appendix}Derivation of the Texture Eq.(\ref{Eq:Mnu-2})}
A set of constraints to have $U_{PMNS}$ as the PMNS unitary matrix, which determines masses and mixing angles \cite{SpecificCP}, is given by
\begin{eqnarray}
&&
{\sin 2\theta_{12} \left( {\lambda _1  - \lambda _2 } \right) + 2\cos 2\theta_{12} X = 0},
\label{Eq:theta12-single}\\
&&
\left( M_{\tau\tau} - M_{\mu\mu}\right)\sin 2\theta_{23}  - 2 M_{\mu\tau}\cos 2\theta_{23}= 2s_{13} e^{ - i\delta } X,
\label{Eq:theta23-single}\\
&&
{\sin 2\theta_{13} \left( {M_{ee}e^{ - i\delta }  - \lambda _3 e^{i\delta } } \right) + 2\cos 2\theta_{13} Y} = 0,
\label{Eq:theta13-single}
\end{eqnarray}
and
\begin{eqnarray}
&&
m_1 e^{ - 2i\beta _1 }  = c_{12}^2 \lambda _1  + s_{12}^2 \lambda _2  - 2c_{12} s_{12} X,
\quad
m_2 e^{ - 2i\beta _2 }  = c_{12}^2 \lambda _1  + s_{12}^2 \lambda _2  + 2c_{12} s_{12} X,
\nonumber\\
&&
m_3 e^{ - 2i\beta _3 }  = \frac{{c_{13}^2 \lambda _3  - s_{13}^2 e^{ - 2i\delta } M_{ee}}}{{c_{13}^2  - s_{13}^2 }}.
\label{Eq:neutrino-masses}
\end{eqnarray}
The mass parameters of $\lambda_{1,2,3}$, $X$ and $Y$ are given by
\begin{eqnarray}
&&
\lambda_1  = c_{13}^2 M_{ee} - 2c_{13} s_{13} e^{i\delta } Y + s_{13}^2 e^{2i\delta }\lambda_3,
\quad
\lambda_2  = c_{23}^2 M_{\mu\mu} + s_{23}^2 M_{\tau\tau} - 2s_{23} c_{23} M_{\mu\tau},
\nonumber\\
&&
\lambda_3  = s_{23}^2 M_{\mu\mu} + c_{23}^2 M_{\tau\tau} + 2s_{23} c_{23} M_{\mu\tau},
\label{Eq:Parameters}\\
&&
X = \frac{c_{23} M_{e\mu} - s_{23} M_{e\tau}}{c_{13}},
\quad
Y = s_{23} M_{e\mu} + c_{23} M_{e\tau}.
\label{Eq:X-Y}
\end{eqnarray}
To have a diagonalized $U^T_{PMNS}M_\nu U_{PMNS}$ yields six constraints.  Three constraints on the diagonal parts give Eq.(\ref{Eq:neutrino-masses}) and other three supply  Eqs.(\ref{Eq:theta12-single})-(\ref{Eq:theta13-single}) for three mixing angles.  It should be noted that each mixing angle is determined twofold by the real part of the equation and its imaginary part, which should be compatible with each other unless the phase is automatically cancelled in both sides of Eqs.(\ref{Eq:theta12-single})-(\ref{Eq:theta13-single}).

The solution to Eq.(\ref{Eq:theta13-single}) for $c_{23}=\sigma s_{23}=1/\sqrt 2$ is found to be:
\begin{eqnarray}
&&
M_{\mu \mu }  = \left( {\frac{1}{{c_{13} \tan 2\theta _{12} }} - \sigma t_{13} e^{ - i\delta } } \right)\sqrt{2}M^{(+)}_{e\mu} + \left( {\frac{{e^{ - i\delta } }}{{\tan 2\theta _{13} }} - \frac{{t_{13} e^{i\delta } }}{2}} \right)\sqrt{2}\sigma M^{(-)}_{e\mu} + \frac{{e^{ - 2i\delta }  + 1}}{2}M_{ee},
\nonumber\\ 
&&
M_{\tau \tau }  = \left( {\frac{1}{{c_{13} \tan 2\theta _{12} }} + \sigma t_{13} e^{ - i\delta } } \right)\sqrt{2}M^{(+)}_{e\mu} + \left( {\frac{{e^{ - i\delta } }}{{\tan 2\theta _{13} }} - \frac{{t_{13} e^{i\delta } }}{2}} \right)\sqrt{2}\sigma M^{(-)}_{e\mu} + \frac{{e^{ - 2i\delta }  + 1}}{2}M_{ee},
\nonumber\\ 
&&
\sigma M_{\mu \tau }  =  - \frac{1}{{c_{13} \tan 2\theta _{12} }}\sqrt{2}M^{(+)}_{e\mu} + \left( {\frac{{e^{ - i\delta } }}{{\tan 2\theta _{13} }} + \frac{{t_{13} e^{i\delta } }}{2}} \right)\sqrt{2}\sigma M^{(-)}_{e\mu} + \frac{{e^{ - 2i\delta }  - 1}}{2}M_{ee},
\label{Eq:Solution}
\end{eqnarray}
from which we have
\begin{eqnarray}
&&
M^{(-)}_{\mu\mu }  =  - \sqrt 2 \sigma t_{13} e^{ - i\delta } M_{e\mu}^{(+)}.
\label{Eq:ExamplesConstraint}
\end{eqnarray}
This is nothing but Eq.(\ref{Eq:theta23-single}). These flavor neutrino masses become identical to those found in Ref.\cite{SpecificCP} if CP violation becomes maximal.  Since we know that the $\mu$-$\tau$ symmetric texture is consistent with the present observation of neutrino oscillations, let us assume that 
\begin{eqnarray}
&&
M_{ee}=a_0,
\quad
M_{e\mu}=b_0+ e^{i\alpha}b^\prime_0,
\quad
M_{e\tau}=-\sigma \left( b_0- e^{i\alpha}b^\prime_0\right),
\label{Eq:Examples}
\end{eqnarray}
where $b^\prime_0$ stands for the minor deviation from the $\mu$-$\tau$ symmetric texture and $\alpha$ is a phase parameter, which gives
\begin{eqnarray}
&&
M^{(+)}_{e\mu} = b_0,
\quad
M^{(-)}_{e\mu} = e^{i\alpha}b^\prime_0.
\label{Eq:mu-tau-related}
\end{eqnarray}
From Eq.(\ref{Eq:ExamplesConstraint}), we find that
\begin{eqnarray}
&&
M^{(-)}_{\mu\mu} = -\sqrt 2 \sigma t_{13} e^{ - i\delta } b_0.
\label{Eq:mu-tau-related2}
\end{eqnarray}
If the $\mu$-$\tau$ symmetry breaking has a common source, it is expected that the phases in $M^{(-)}_{e\mu}$ and $M^{(-)}_{\mu\mu}$ are the same:
\begin{eqnarray}
&&
\alpha= - \delta,
\label{Eq:mu-tau-related3}
\end{eqnarray}
and that their strength takes the similar magnitude:
\begin{eqnarray}
&&
b^\prime_0\sim t_{13}b_0,
\label{Eq:mu-tau-related4}
\end{eqnarray}
as well.

Collecting these results, we find that Eq.(\ref{Eq:Solution}) becomes
\begin{eqnarray}
&&
M_{\mu \mu }  = \left( {\frac{1}{{c_{13} \tan 2\theta _{12} }} - \sigma t_{13} e^{ - i\delta } } \right){\sqrt 2 }b_0 + \left( {\frac{{e^{ - 2i\delta } }}{{\tan 2\theta _{13} }} - \frac{t_{13} }{2}} \right){\sqrt 2 }\sigma b^\prime_0  + \frac{{e^{ - 2i\delta }  + 1}}{2}a_0,
\nonumber\\ 
&&
M_{\tau \tau }  = \left( {\frac{1}{{c_{13} \tan 2\theta _{12} }} + \sigma t_{13} e^{ - i\delta } } \right){\sqrt 2 }b_0 + \left( {\frac{{e^{ - 2i\delta } }}{{\tan 2\theta _{13} }} - \frac{t_{13} }{2}} \right){\sqrt 2 }\sigma b^\prime_0  + \frac{{e^{ - 2i\delta }  + 1}}{2}a_0,
\nonumber\\ 
&&
\sigma M_{\mu \tau }  =  - \frac{1}{{c_{13} \tan 2\theta _{12} }}{\sqrt 2 }b_0 + \left( {\frac{{e^{ - 2i\delta } }}{{\tan 2\theta _{13} }} + \frac{t_{13} }{2}} \right){\sqrt 2 }\sigma b^\prime_0 + \frac{{e^{ - 2i\delta }  - 1}}{2}a_0.
\label{Eq:Examples2}
\end{eqnarray}
From these expressions, we finally reach the following parameterization:
\begin{eqnarray}
&&
M_{\mu \mu } = d_0+e^{ - i\delta }d^\prime_0 + e^{ - 2i\delta }d_1 ,
\quad
M_{\tau \tau } = d_0-e^{ - i\delta }d^\prime_0 + e^{ - 2i\delta }d_1,
\quad
M_{\mu \tau } = \sigma \left( { - d_0  + e^{ - 2i\delta }d_1  } \right),
\label{Eq:Examples3}
\end{eqnarray}
which is the mass texture of Eq.(\ref{Eq:Mnu-2}).  By comparing Eq.(\ref{Eq:Examples3}) with Eq.(\ref{Eq:Examples2}), we also find that
\begin{eqnarray}
&&
\tan 2\theta _{12} \approx 2\sqrt 2 \frac{{b_0 }}{{2d_0  - a_0 }},
\quad
\tan 2\theta _{13}  = 2\sqrt 2 \sigma \frac{{b^\prime_0 }}{{2d_1  - a_0 }},
\label{Eq:Results}
\end{eqnarray}
where the approximation of $\sin^2\theta_{13}\approx 0$ is used to show $\tan 2\theta _{12}$ and 
\begin{eqnarray}
t_{13}  =  - \sigma \frac{d^\prime_0 }{{\sqrt 2}b_0 },
\label{Eq:Results2}
\end{eqnarray}
from Eq.(\ref{Eq:mu-tau-related2}).  The mass parameter $d^\prime_0$ is so determined that two expressions for $\theta_{13}$ are consistent with each other.  From Eqs.(\ref{Eq:Results}) and (\ref{Eq:Results2}), we find that the $\mu$-$\tau$ symmetry breaking masses of $b^\prime_0$ and $d^\prime_0$ can be parameterized by a common parameter $\varepsilon$ satisfying $\vert\varepsilon\vert\ll 1$ to be $b^\prime_0\propto \varepsilon$ and $d^\prime_0\propto \varepsilon$.


\noindent

\noindent

\end{document}